\documentclass[12pt]{iopart}

\usepackage{graphicx}

\begin{document}

\title[Surface stresses on a thin shell surrounding a traversable wormhole]
{Surface stresses on a thin shell surrounding a traversable
wormhole}

\author{Francisco S. N. Lobo\footnote[1]{flobo@cosmo.fis.fc.ul.pt} }

\address{Centro de Astronomia
e Astrof\'{\i}sica da Universidade de Lisboa,\\
Campo Grande, Ed. C8 1749-016 Lisboa, Portugal}


\begin{abstract}

We match an interior solution of a spherically symmetric
traversable wormhole to a unique exterior vacuum solution, with a
generic cosmological constant, at a junction interface, and the
surface stresses on the thin shell are deduced. In the spirit of
minimizing the usage of exotic matter we determine regions in
which the weak and null energy conditions are satisfied on the
junction surface. The characteristics and several physical
properties of the surface stresses are explored, namely, regions
where the sign of the tangential surface pressure is positive and
negative (surface tension) are determined. This is done by
expressing the tangential surface pressure as a function of
several parameters, namely, that of the matching radius, the
redshift parameter, the surface energy density and of the generic
cosmological constant. An equation governing the behavior of the
radial pressure across the junction surface is also deduced.

\end{abstract}

\pacs{04.20.-q, 04.20.Jb, 04.40.-b}

\maketitle



\section{Introduction}

Interest in traversable wormholes, as hypothetical shortcuts in
spacetime, has been rekindled by the classical paper by Morris and
Thorne \cite{Morris}. The subject has also served to stimulate
research in several branches, namely the energy condition
violations \cite{Visser,VisserEC}, time machines and the
associated difficulties in causality violation
\cite{Visser,timetravel}, and superluminal travel
\cite{superluminal}, amongst others.

As the violation of the energy conditions is a particularly
problematic issue \cite{VisserEC}, it is useful to minimize the
usage of exotic matter \cite{VKD,Hochberg}. Recently, Visser {\it
et al} \cite{VKD}, by introducing the notion of the ``volume
integral quantifier'', found specific examples of spacetime
geometries containing wormholes that are supported by arbitrarily
small quantities of averaged null energy condition violating
matter, although the null energy and averaged null energy
conditions are always violated for wormhole spacetimes. Another
elegant way of minimizing the usage of exotic matter is to
construct a simple class of wormhole solutions using the cut and
paste technique \cite{Visser,visser1}, in which the exotic matter
is concentrated at the wormhole throat. The surface stresses of
the exotic matter were determined by invoking the Darmois-Israel
formalism \cite{Israel}. These thin-shell wormholes are extremely
useful as one may apply a stability analysis for the dynamical
cases, either by choosing specific surface equations of state
\cite{eqstate}, or by considering a linearized stability analysis
around a static solution \cite{linear}, in which a parametrization
of the stability of equilibrium is defined, so that one does not
have to specify a surface equation of state.

In fact, the Darmois-Israel thin shell formalism, a fundamental
tool in Classical General Relativity, has found extensive
applications in the literature, ranging from the gravitational
collapse \cite{gravcollapse}, the evolution of bubbles and domain
walls in cosmological settings \cite{domainwall}, shells around
black hole solutions \cite{FHK}, black holes as possible sources
of closed and semiclosed universes \cite{FMM}, and matchings of
cosmological solutions \cite{SakaiMaeda}, to the Randall-Sundrum
brane world scenario \cite{Randall,brane}, where our universe is
viewed as a domain wall in five dimensional anti-de Sitter space.
An interesting application to wormhole physics was done in
\cite{Frolov}, where Frolov and Novikov demonstrated that if one
of the wormhole mouths is inserted in an external gravitational
field, for instance, a thin shell is placed around one of the
traversable wormhole mouths, the Killing vector field is
well-defined locally but does not exist globally, enabling the
transformation of the wormhole into a time machine. More recently,
inspired in the {\it gravastar} ({\it gra}vitational {\it va}cuum
{\it star}) picture, an alternative to black holes, developed by
Mazur and Mottola \cite{Mazur}, where there is a phase transition
at/near $2M$, and the interior Schwarzschild solution is replaced
by a segment of the de Sitter space, Visser and Waltshire
\cite{VissWalt}, using the thin shell formalism, constructed a
model sharing the key features of the Mazur-Mottola scenario, and
analyzed the dynamic stability of the configuration against radial
perturbations.

As an alternative to the thin-shell wormhole \cite{visser1}, one
may also consider that the exotic matter, threading an interior
wormhole spacetime, is distributed from the throat to a radius
$a$, where the solution is matched to an exterior vacuum
spacetime. Several simple cases were analyzed in \cite{Morris},
but one may invoke the Darmois-Israel formalism to consider a
broader class of solutions. Thus, the thin shell confines the
exotic matter to a finite region, with a delta-function
distribution of the stress-energy tensor on the junction surface.
One of the motivations of this construction, apart from
constraining the exotic matter threading the interior wormhole to
(arbitrarily small) finite regions, resides in determining domains
in which the surface stresses of the thin shell obey the energy
conditions, in order to further minimize the usage of exotic
matter.

In the present work, a thin shell surrounding a spherically
symmetric traversable wormhole, with a generic cosmological
constant, is analyzed. A general class of wormhole geometries with
a cosmological constant and junction conditions was analyzed by
DeBenedictis and Das~\cite{DeDas1}, and further explored in higher
dimensions~\cite{DeDas2}. It is of interest to study a positive
cosmological constant, as the inflationary phase of the
ultra-early universe demands it, and in addition, recent
astronomical observations point to $\Lambda
>0$. On the other hand, a negative cosmological constant is the
vacuum state for extended theories of gravitation, such as
supergravity and superstring theories. We generalize and
systematize the particular case of a matching with a constant
redshift function and a null surface energy density on the
junction boundary, studied in \cite{LLQ}. A similar analysis for
the plane symmetric case, with a negative cosmological constant,
is done in \cite{LL}. The plane symmetric traversable wormhole is
a natural extension of the topological black hole solutions found
by Lemos \cite{lemos}, upon addition of exotic matter. These plane
symmetric wormholes may be viewed as domain walls connecting
different universes, having planar topology, and upon
compactification of one or two coordinates, cylindrical topology
or toroidal topology, respectively.

The plan of this paper is as follows: In sections 2 and 3, we
present the interior wormhole solution and the unique exterior
vacuum solution, respectively. In section 4, we present the
junction conditions, by deducing the surface stresses; we find
specific regions where the energy conditions at the junction are
obeyed; we also analyze the physical properties and
characteristics of the surface stresses, namely, we find domains
where the tangential surface pressure is positive or negative
(surface tension); and finally, we deduce an expression governing
the behavior of the radial pressure across the junction boundary.
Finally, in section 5, we conclude.


\section{Interior solution}

The spacetime metric representing a spherically symmetric and
static wormhole is given by
\begin{equation}
ds^2=-e ^{2\Phi(r)}\,dt^2+\frac{dr^2}{1- b(r)/r}+r^2 \,(d\theta
^2+\sin ^2{\theta} \, d\phi ^2) \label{metricwormhole}\,,
\end{equation}
where $\Phi(r)$ and $b(r)$ are arbitrary functions of the radial
coordinate, $r$. $\Phi(r)$ is denoted as the redshift function,
for it is related to the gravitational redshift; $b(r)$ is called
the form function, because as can be shown by embedding diagrams,
it determines the shape of the wormhole \cite{Morris}. The radial
coordinate has a range that increases from a minimum value at
$r_0$, corresponding to the wormhole throat, to $a$, where the
interior spacetime will be joined to an exterior vacuum solution.

Using the Einstein field equation with a non-vanishing
cosmological constant, $G_{\hat{\mu}\hat{\nu}}+\Lambda
\eta_{\hat{\mu}\hat{\nu}}=8\pi \,T_{\hat{\mu}\hat{\nu}}$, in an
orthonormal reference frame, (with $c=G=1$) we obtain the
following stress-energy scenario
\begin{eqnarray}
\rho(r)&=&\frac{1}{8\pi} \left(\frac{b'}{r^2}-\Lambda \right)  \label{rhoWHlambda}\,,\\
\tau (r)&=&\frac{1}{8\pi} \left[\frac{b}{r^3}-2
\left(1-\frac{b}{r}
\right) \frac{\Phi'}{r}- \Lambda \right]  \label{tauWHlambda}\,,\\
p(r)&=&\frac{1}{8\pi} \Bigg \{\left(1-\frac{b}{r}\right)\Bigg[\Phi
''+ (\Phi')^2- \frac{b'r-b}{2r(r-b)}\Phi'
       \nonumber     \\
&&-\frac{b'r-b}{2r^2(r-b)}+\frac{\Phi'}{r} \Bigg] + \Lambda \Bigg
\} \label{pressureWHlambda}\,,
\end{eqnarray}
in which $\rho(r)$ is the energy density; $\tau (r)$ is the radial
tension; $p(r)$ is the pressure measured in the lateral
directions, orthogonal to the radial direction.

One may readily verify that the null energy condition (NEC) is
violated at the wormhole throat. The NEC states that
$T_{\mu\nu}k^{\mu}k^{\nu}\geq0$, where $k^{\mu}$ is a null vector.
In the orthonormal frame, $k^{\hat{\mu}}=(1,1,0,0)$, we have
\begin{equation}\label{NECthroat}
T_{\hat{\mu}\hat{\nu}}k^{\hat{\mu}}k^{\hat{\nu}}=\rho(r)-\tau(r)=
\frac{1}{8\pi}\,\left[\frac{b'r-b}{r^3}+
2\left(1-\frac{b}{r}\right) \frac{\Phi '}{r} \right]  .
\end{equation}
Due to the flaring out condition of the throat deduced from the
mathematics of embedding, i.e., $(b-b'r)/b^2>0$
\cite{Morris,Visser,LLQ}, we verify that at the throat
$b(r_0)=r=r_0$, and due to the finiteness of $\Phi(r)$, from
equation (\ref{NECthroat}) we have
$T_{\hat{\mu}\hat{\nu}}k^{\hat{\mu}}k^{\hat{\nu}}<0$. Matter that
violates the NEC is denoted as exotic matter. The exotic matter
threading the wormhole extends from the throat at $r_0$ to the
junction boundary situated at $a$, where the interior solution is
matched to an exterior vacuum spacetime.

More recently, Visser {\it et al}~\cite{VKD}, noting the fact that
the energy conditions do not actually quantify the ``total
amount'' of energy condition violating matter, developed a
suitable measure for quantifying this notion by introducing a
``volume integral quantifier''. This notion amounts to calculating
the definite integrals $\int T_{\mu\nu}U^\mu U^\nu \,dV$ and $\int
T_{\mu\nu}k^\mu k^\nu \,dV$, and the amount of violation is
defined as the extent to which these integrals become negative.
(Recently, by using the ``volume integral quantifier'',
fundamental limitations on ``warp drive'' spacetimes in the weak
field limit were also found~\cite{LVwarp}). For instance, the
integral which provides information about the ``total amount'' of
averaged null energy condition (ANEC) violating matter in the
spacetime is given by (see \cite{VKD} for details)
\begin{eqnarray}\label{vol:int}
\int \left[\rho(r) -\tau(r)\right]\; dV= -\int_{r_0}^\infty
\left(1-b' \right) \;\left[\ln \left(\frac{e^{\Phi}}{1-b/r}
\right) \right] \;dr \,.
\end{eqnarray}
Now considering specific choices for the form function and
matching the interior solution to an exterior solution at $a$,
Visser {\it et al} found specific examples of spacetime geometries
containing wormholes that are supported by arbitrarily small
quantities of ANEC violating matter, although the null energy and
averaged null energy conditions are always violated for wormhole
spacetimes.

\section{Exterior solution}

The exterior solution is given by
\begin{eqnarray}
ds^2&=&-g(r)\,dt^2+g(r)^{-1}dr^2+r^2(d\theta ^2+\sin ^2{\theta}\,
d\phi ^2) \label{metricvacuumlambda}
\end{eqnarray}
with
\begin{equation}
g(r)=1-\frac{2M}{r}-\frac{\Lambda}{3}r^2  \,.
\end{equation}
If $\Lambda >0$, the solution is denoted by the Schwarzschild-de
Sitter spacetime. For $\Lambda <0$, we have the Schwarzschild-anti
de Sitter spacetime, and of course the specific case of $\Lambda
=0$ is reduced to the Schwarzschild solution, with a black hole
event horizon at $r_b=2M$. Note that the metric $(7)$ is not
asymptotically flat as $r \rightarrow \infty$. Rather, it is
asymptotically de Sitter, if $\Lambda >0$, or asymptotically
anti-de Sitter, if $\Lambda <0$.

Consider the Schwarzschild-de Sitter spacetime, $\Lambda
>0$. If $0<9\Lambda M^2<1$, the factor $g(r)=(1-2M/r-\Lambda r^2/3)$
possesses two positive real roots, $r_b$ and $r_c$, corresponding
to the black hole and the cosmological event horizons,
respectively, given by
\begin{eqnarray}
r_b&=&2 \Lambda ^{-1/2} \, \cos(\alpha/3)    \label{root1}  \,, \\
r_c&=&2 \Lambda ^{-1/2} \, \cos(\alpha/3+4\pi/3)    \label{root2}
\,,
\end{eqnarray}
where $\cos \alpha \equiv -3M \Lambda^{1/2}$, with $\pi < \alpha
<3\pi/2$. In this domain we have $2M<r_b<3M$ and $r_c>3M$.

For the Schwarzschild-anti de Sitter metric, with $\Lambda <0$,
the factor $g(r)=\left(1-2M/r+|\Lambda |r^2/3 \right)$ has only
one real positive root, $r_b$, given by
\begin{eqnarray}
r_b=\left(\frac{3M}{|\Lambda|}\right)^{1/3}\Bigg(\sqrt[3]
{1+\sqrt{1+\frac{1}{9|\Lambda|M^2}}} +\sqrt[3]{1-\sqrt{1+
\frac{1}{9|\Lambda|M^2}}}\;\Bigg) \label{adsbhole} ,
\end{eqnarray}
corresponding to a black hole event horizon, with $0<r_b<2M$.

\section{Junction conditions}

\subsection{The surface stresses}

We shall match the interior solution, equation
(\ref{metricwormhole}), to the exterior vacuum solution, equation
(\ref{metricvacuumlambda}), at a junction surface, $\Sigma$.
Consider the junction surface $\Sigma$ as a timelike hypersurface
defined by the parametric equation of the form
$f(x^{\mu}(\xi^i))=0$. $\xi^i=(\tau,\theta,\phi)$ are the
intrinsic coordinates on $\Sigma$, where $\tau$ is the proper time
on the hypersurface. The three basis vectors tangent to $\Sigma$
are given by $e_{(i)}=\partial/\partial \xi^i$, with the following
components $e^{\mu}_{(i)}=\partial x^{\mu}/\partial \xi^i$. The
induced metric on the junction surface is then provided by the
scalar product $g_{ij}=e_{(i)}\cdot e_{(j)}=g_{\mu
\nu}e^{\mu}_{(i)}e^{\nu}_{(j)}$. Thus, the intrinsic metric to
$\Sigma$ is given by
\begin{equation}
ds^2_{\Sigma}=-d\tau^2 + a^2 \,(d\theta ^2+\sin
^2{\theta}\,d\phi^2)  \,.
\end{equation}
Note that the junction surface, $r=a$, is situated outside the
event horizon, i.e., $a>r_b$, to avoid a black hole solution.

The unit normal $4-$vector, $n^{\mu}$, to $\Sigma$ is defined as
\begin{equation}
n_{\mu}=\pm \,\left |g^{\alpha \beta}\,\frac{\partial f}{\partial
x ^{\alpha}} \, \frac{\partial f}{\partial x ^{\beta}}\right
|^{-1/2}\;\frac{\partial f}{\partial x^{\mu}}\,,
\end{equation}
with $n_{\mu}\,n^{\mu}=+1$ and $n_{\mu}e^{\mu}_{(i)}=0$. The
Israel formalism requires that the normals point from the interior
spacetime to the exterior spacetime. The extrinsic curvature, or
the second fundamental form, is defined as
$K_{ij}=n_{\mu;\nu}e^{\mu}_{(i)}e^{\nu}_{(j)}$. Differentiating
$n_{\mu}e^{\mu}_{(i)}=0$ with respect to $\xi^j$, we have
$n_{\mu}\frac{\partial ^2 x^{\mu}}{\partial \xi^i \, \partial
\xi^j}=-n_{\mu,\nu}\,\frac{\partial x^{\mu}}{\partial
\xi^i}\frac{\partial x^{\nu}}{\partial \xi^j}$, so that the
extrinsic curvature is finally given by
\begin{eqnarray}\label{extrinsiccurv}
K_{ij}^{\pm}=-n_{\mu} \left(\frac{\partial ^2 x^{\mu}}{\partial
\xi ^{i}\,\partial \xi ^{j}}+\Gamma ^{\mu \pm}_{\;\;\alpha
\beta}\;\frac{\partial x^{\alpha}}{\partial \xi ^{i}} \,
\frac{\partial x^{\beta}}{\partial \xi ^{j}} \right) \,,
\end{eqnarray}
where the $(\pm)$ superscripts correspond to the exterior and
interior spacetimes, respectively. Note that, in general, $K_{ij}$
is not continuous across $\Sigma$, so that for notational
convenience, the discontinuity in the extrinsic curvature is
defined as $\kappa_{ij}=K_{ij}^{+}-K_{ij}^{-}$.

The Einstein equations may be written in the following form,
$S^{i}_{\;j}=-\frac{1}{8\pi}\,(\kappa ^{i}_{\;j}-\delta
^{i}_{\;j}\kappa ^{k}_{\;k})$, denoted as the Lanczos equations,
where $S^{i}_{\;j}$ is the surface stress-energy tensor on
$\Sigma$. Considerable simplifications occur due to spherical
symmetry, namely $\kappa ^{i}_{\;j}={\rm diag} \left(\kappa
^{\tau}_{\;\tau},\kappa ^{\theta}_{\;\theta},\kappa
^{\theta}_{\;\theta}\right)$. The surface stress-energy tensor may
be written in terms of the surface energy density, $\sigma$, and
the surface pressure, ${\cal P}$, as $S^{i}_{\;j}={\rm
diag}(-\sigma,{\cal P},{\cal P})$. The Lanczos equations then
reduce to
\begin{eqnarray}
\sigma &=&-\frac{1}{4\pi}\,\kappa ^{\theta}_{\;\theta} \,,\label{sigma} \\
{\cal P} &=&\frac{1}{8\pi}(\kappa ^{\tau}_{\;\tau}+\kappa
^{\theta}_{\;\theta}) \,, \label{surfacepressure}
\end{eqnarray}
which simplifies the determination of the surface stress-energy
tensor to that of the calculation of the non-trivial components of
the extrinsic curvature. Thus, using equation
(\ref{extrinsiccurv}), the latter are given by
\begin{eqnarray}
K ^{\tau \;+}_{\;\;\tau}&=&\frac{\frac{M}{a^2}-
\frac{\Lambda}{3}a}{\sqrt{1-\frac{2M}{a}-\frac{\Lambda}{3}a^2}}
\;,  \label{Kplustautau}\\
K ^{\tau \;-}_{\;\;\tau}&=&\Phi'(a)\sqrt{1-\frac{b(a)}{a}}  \;,
\label{Kminustautau}
\end{eqnarray}
and
\begin{eqnarray}
K ^{\theta \;+}_{\;\;\theta}&=&\frac{1}{a}\sqrt{1-\frac{2M}{a}-
\frac{\Lambda}{3}a^2}\;,  \label{Kplustheta}\\
K ^{\theta \;-}_{\;\;\theta}&=&\frac{1}{a}\sqrt{1-\frac{b(a)}{a}}
\;.  \label{Kminustheta}
\end{eqnarray}

\begin{figure}[h]
  \centering
  \includegraphics[width=3.2in]{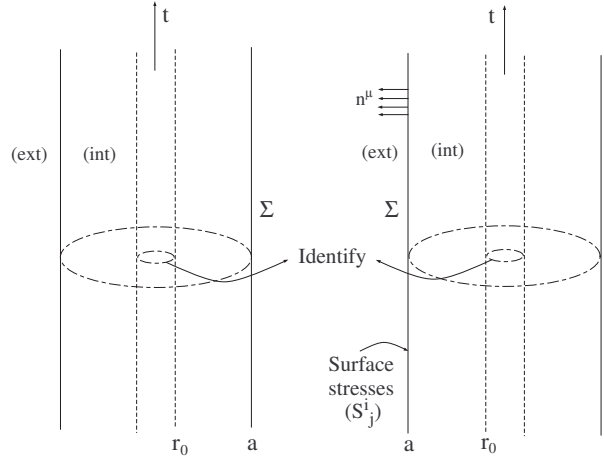}
  \caption{Two copies of static timelike hypersurfaces, $\Sigma$, embedded in
  asymptotic regions, separating an interior wormhole solution from
  an exterior vacuum spacetime. Both copies are identified at the wormhole
  throat, $r_0$. The surface stresses reside on $\Sigma$, and members of the
  normal vector field, $n^{\mu}$, are shown.}\label{fig1}
\end{figure}

The Einstein equations,
equations(\ref{sigma})-(\ref{surfacepressure}), with the extrinsic
curvatures, equations(\ref{Kplustautau})-(\ref{Kminustheta}), then
provide us with the following expressions
\begin{eqnarray}
\sigma&=&-\frac{1}{4\pi a} \left(\sqrt{1-\frac{2M}{a}-
\frac{\Lambda}{3}a^2}- \sqrt{1-\frac{b(a)}{a}} \, \right)
    \label{surfenergy}   ,\\
{\cal P}&=&\frac{1}{8\pi a} \left(\frac{1-\frac{M}{a}
-\frac{2\Lambda}{3}a^2}{\sqrt{1-\frac{2M}{a}-\frac{\Lambda}{3}a^2}}-
\zeta \, \sqrt{1-\frac{b(a)}{a}} \, \right)
    \label{surfpressure}    ,
\end{eqnarray}
with $\zeta=1+a\Phi'(a)$. If the surface stress-energy terms are
null, the junction is denoted as a boundary surface. If surface
stress terms are present, the junction is called a thin shell,
which is represented in figure \ref{fig1}.

The surface mass of the thin shell is given by $M_{\rm shell}=4\pi
a^2\sigma$ or
\begin{eqnarray}\label{shellmass}
M_{\rm shell}=a\left(\sqrt{1-\frac{b(a)}{a}}-\sqrt{1-\frac{2M}{a}-
\frac{\Lambda}{3}a^2} \, \right)  .
\end{eqnarray}
One may interpret $M$ as the total mass of the system, in this
case being the total mass of the wormhole in one asymptotic
region. Thus, solving equation (\ref{shellmass}) for $M$, we
finally have
\begin{equation}\label{totalmass}
M=\frac{b(a)}{2}+M_{\rm
shell}\left(\sqrt{1-\frac{b(a)}{a}}-\frac{M_{\rm
shell}}{2a}\right)-\frac{\Lambda}{6}a^3   \,.
\end{equation}


It is interesting to find some estimates of the surface stresses,
for a specific choice of the form function, $b(r)$. Considering
dimensionless parameters, equations
(\ref{surfenergy})-(\ref{surfpressure}), for the Schwarzschild-de
Sitter solution, take the following form
\begin{eqnarray}
\mu&=&\xi \left(\sqrt{1-\xi\,\bar{b}(\xi)}-\sqrt{1-\xi-
\frac{4\beta}{27\xi^2}} \; \right)
    \label{mu}   ,\\
\Pi&=&\xi \left(\frac{1-\frac{\xi}{2}
-\frac{8\beta}{27\xi^2}}{\sqrt{1-\xi-\frac{4\beta}{27\xi^2}}}-
\zeta \, \sqrt{1-\xi\bar{b}(\xi)} \, \right)
    \label{Pi}    ,
\end{eqnarray}
with the following definitions: $\xi=2M/a$, $\beta=9\Lambda M^2$,
$\bar{b}(\xi)=b(a)/(2M)$, $\mu=8\pi M\sigma$ and $\Pi=16\pi M{\cal
P}$. In the analysis that follows we shall assume that $M$ is
positive, $M>0$. The surface stresses for the Schwarzschild
solution are obtained by setting $\Lambda=0$, i.e., $\beta=0$.

\begin{figure}[h]
  \centering
  \includegraphics[width=2.4in]{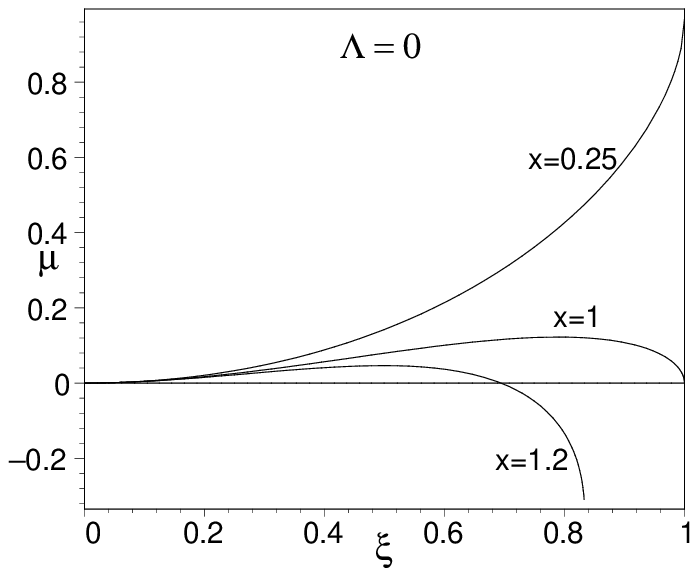}
  \hspace{0.4in}
  \includegraphics[width=2.4in]{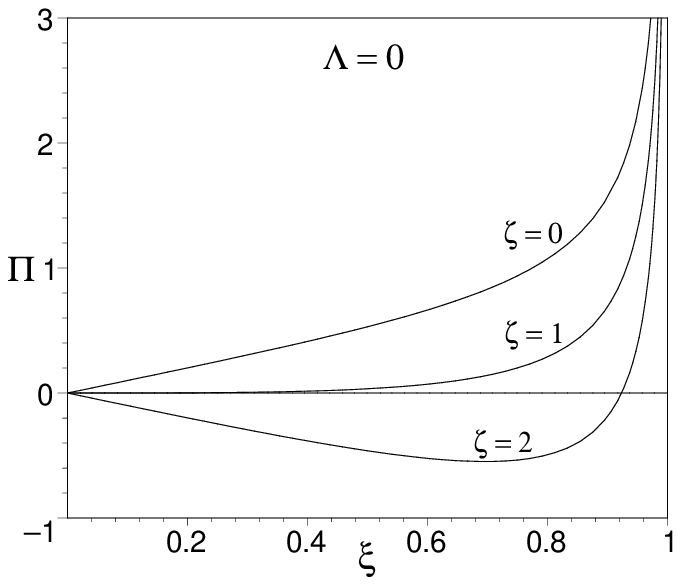}
  \caption{The Schwarzschild spacetime,
  $\Lambda=0$, with the particular choice of $b(r)=r_0^2/r$.
  Plots of the surface energy density and surface pressure, are
  given in terms of dimensionless parameters, i.e., $\mu=8\pi M\sigma$
  and $\Pi=16\pi M{\cal P}$, respectively. We have also defined $\xi=2M/a$
  and $x=r_0/2M$. For $x\leq 1$, a non-negative
  surface energy density, $\mu\geq 0$, is obtained;
  for $x>1$, a negative surface energy
  density is obtained for large values of $\xi$. In the $\Pi$ plot
  (with $x=0.25$),
  for $\zeta\leq 1$, we have a surface pressure, $\Pi>0$,
  while for $\zeta>1$, surface tensions are obtained for low values of
  $\xi$. See text for details.
  }\label{fig2}
\end{figure}

Considering a specific choice of a form function, for instance,
$b(r)=r_0^2/r$, $\mu$ and $\Pi$ are represented in figure
\ref{fig2}, for the Schwarzschild solution. We have defined
$x=r_0/(2M)$, so that for $x>1$ ($r_0>2M$), the range of $\xi$ is
given by $0<\xi<1/x$. For the $\mu$ plot, we have considered the
cases of $x=0.25$, $x=1$ and $x=1.2$, i.e., $r_0=M/2$, $r_0=2M$
and $r_0=2.4M$, respectively. We verify that for $x\leq 1$, one
always obtains a non-negative surface energy density, $\mu\geq 0$,
whilst for $x>1$ a negative surface energy density is obtained for
large values of $\xi$ (for $x=1.2$, the range of $\xi$ is
restricted to $0<\xi<0.833$). In relationship to the $\Pi$ plot,
we have only considered $x=0.25$ ($r_0=M/2$), as the qualitative
behaviour for arbitrary $x$ is similar to the one presented. For
$\zeta\leq 1$, a surface pressure, $\Pi>0$, is obtained, while for
$\zeta>1$, surface tensions are obtained for low values of $\xi$
(high $a$).

\begin{figure}[h]
  \centering
  \includegraphics[width=2.4in]{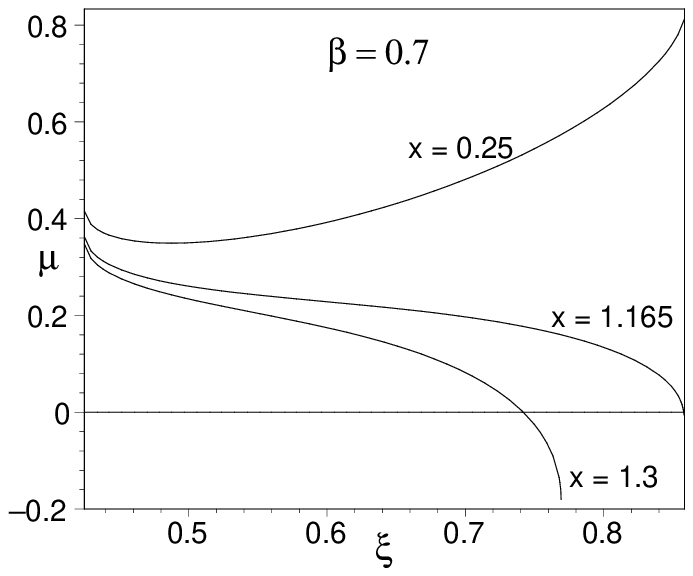}
  \hspace{0.4in}
  \includegraphics[width=2.4in]{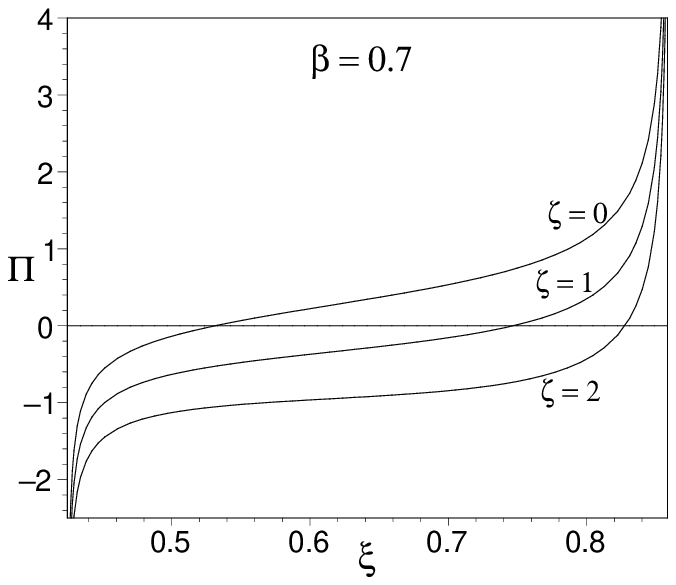}
  \caption{Plots of $\mu$ and $\Pi$
  for the Schwarzschild-de Sitter spacetime, $\Lambda>0$, with the particular
  choice of $b(r)=r_0^2/r$ and $\beta=9\Lambda M^2=0.7$.
  The qualitative behaviour for arbitrary $\beta$ is similar to the one
  presented. The black hole and cosmological horizons are given by
  $r_b \simeq 2.33\,M$ and $r_c =4.71\,M$, respectively. For $x\leq 1.165$,
  i.e., $r_0\leq r_b$, $\mu$ is non-negative; for $x> 1.165$, i.e., $r_0>r_b$,
  a negative surface energy density is obtained for high values of $\xi$.
  In the $\Pi$ graph, we have set $x=0.25$, i.e., $r_0=M/2$, with $\zeta=0$, $\zeta=1$
  and $\zeta=2$, respectively.
  For high values of $\xi$ a surface pressure is needed to hold the
  structure against collapse, and for low $\xi$ a surface tension is
  needed to hold the wormhole structure against expansion.
  See text for details.}\label{fig3}
\end{figure}

For the Schwarzschild-de Sitter case, we have considered
$\beta=9\Lambda M^2=0.7$. The qualitative behaviour for an
arbitrary $\beta$ is similar to the analysis presented in figure
\ref{fig3}. For $\beta=0.7$ the black hole and cosmological
horizons are given by $r_b \simeq 2.33\,M$ and $r_c =4.71\,M$,
respectively. Thus, only the interval $0.425 < \xi < 0.858$ is
taken into account, as shown in the range of the respective plots.
For the $\mu$ plot, we have chosen the values of $x=0.25$,
$x=1.165$ and $x=1.3$, i.e., $r_0=M/2$, $r_0=2.33M=r_b$ and
$r_0=2.6M$, respectively. For $x\leq 1.165$, i.e., $r_0\leq r_b$,
$\mu$ is non-negative; for $x> 1.165$, i.e., $r_0>r_b$, a negative
surface energy density is obtained for high values of $\xi$.
Recall that for $r_0>r_b$ the upper limit of $\xi$ is restricted
by $\xi<1/x=2M/r_0$, so that for the particular case of
$\beta=0.7$ and $x=1.3$, we have the range $0.425 < \xi < 0.769$.
The qualitative behaviour of $\Pi$ is transparent from figure
\ref{fig3}, where we have considered the cases of $\zeta=0$,
$\zeta=1$ and $\zeta=2$, respectively, and $x=0.25$, i.e.,
$r_0=M/2$. The qualitative behaviour of $\Pi$ for arbitrary $x$ is
similiar the one presented. For high values of $\xi$ a surface
pressure is needed to hold the structure against collapse, and for
low $\xi$ a surface tension is needed to hold the wormhole
structure against expansion.

One can do an identical analysis for the anti-de Sitter solution,
as done in the previous case, however this shall not be attempted
here. We shall analyze below the physical properties and
characteristics of the surface stresses for generic wormholes,
i.e., generic $b(r)$ and $\Phi(r)$.

\subsection{The redshift parameter, $\zeta$}

To find a physical significance of the parameter $\zeta$, consider
the redshift function given by $\Phi(r)=kr^{\alpha}$, with
$\alpha, k\in {\bf R}$. Thus, from the definition of
$\zeta=1+a\Phi'(a)$, the redshift function, in terms of $\zeta$,
takes the following form
\begin{equation}\label{redshift}
\Phi(r)=\frac{\zeta-1}{\alpha}\,\left(\frac{r}{a}\right)^{\alpha}
,
\end{equation}
with $\alpha \neq 0$, reducing to $\Phi(a)=(\zeta-1)/\alpha$ at
the junction surface. Thus, $\zeta$ may also be defined as
$\zeta=1+\alpha \Phi(a)$, and one sees that it is related to the
gravitational redshift, through the redshift function, at the
junction interface. Therefore, one may denote $\zeta$ as the
redshift parameter. The proper time on the thin shell is given by
$d\tau=e^{\Phi(a)}dt=e^{(\zeta-1)/\alpha}dt$.

The case of $\alpha=0$ corresponds to the constant redshift
function, so that $\zeta=1$. If $\alpha<0$, then $\Phi(r)$ is
finite throughout spacetime and in the absence of an exterior
solution we have $\lim_{r\rightarrow \infty} \Phi(r)\rightarrow
0$. As we are considering a matching of an interior solution with
an exterior solution at $a$, then it is also possible to consider
the $\alpha>0$ case, imposing that $\Phi(r)$ is finite in the
interval $r_0 \leq r \leq a$.

For the particular case of $\alpha<0$, $\zeta$ is given by
$\zeta=1-|\alpha| ka^{\alpha}$. We verify that if $k>0$, then
$\zeta <1$ and $\zeta$ takes negative values if
$|\alpha|ka^{\alpha}>1$. Thus, the proper time on the thin shell
is given by $d\tau=e^{|\zeta-1|/|\alpha|}dt$, and the condition
$d\tau>dt$ is verified, i.e., proper time on the thin shell flows
faster than the coordinate time, $t$.

If $k<0$, then $\zeta=1+|\alpha||k|a^{\alpha}$ such that $\zeta
>1$. $\zeta$ may take large positive values if $\alpha
ka^{\alpha}$ is sufficiently large. Thus,
$d\tau=e^{-(\zeta-1)/|\alpha|}dt$, with $d\tau<dt$, i.e., proper
time on the thin shell flows slower than the coordinate time, $t$.

For the $\alpha>0$ case we are only interested in the range $r_0
\leq r \leq a$, so that we need to impose that $\Phi(r)$ is finite
in the respective domain. We have $\zeta=1+\alpha ka^{\alpha}$, so
that if $k>0$ then $\zeta
>1$. Therefore $d\tau=e^{(\zeta-1)/\alpha}dt$, implying $d\tau>dt$.

If $k<0$, then from $\zeta=1-\alpha|k|a^{\alpha}$ we have $\zeta
<1$. Thus $d\tau=e^{-|\zeta-1|/\alpha}dt$, and therefore
$d\tau<dt$, i.e., proper time on the thin shell ticks slower than
the coordinate time, $t$.

\subsection{Energy conditions on the junction surface}

The junction surface may serve to confine the interior wormhole
exotic matter to a finite region, which in principle may be made
arbitrarily small. Using the notion of the ``volume integral
quantifier'', equation (\ref{vol:int}), interior wormhole
solutions supported by arbitrarily small quantities of ANEC
violating matter were found \cite{VKD}, although the NEC and WEC
are always violated for wormhole spacetimes. In the spirit of
minimizing the usage of the exotic matter, one may find regions
where the surface stress-energy tensor obeys the energy conditions
at the junction, $\Sigma$ \cite{Visser,hawkingellis}. Thus, we
construct models of spherically symmetric traversable wormholes,
where the ANEC violating matter is confined to a region $r_0\leq
r<a$ (which can be made arbitrarily small by taking the limit $a
\to r_0$), and the thin shell comprises {\it quasi-normal} matter.
Here we consider that {\it quasi-normal} means matter that
satisfies the WEC and NEC (see \cite{VKD} for similar
definitions).

We shall only consider the weak energy condition (WEC) and the
null energy condition (NEC). The WEC implies $\sigma \geq 0$ and
$\sigma + {\cal P} \geq 0$, and by continuity implies the null
energy condition (NEC), $\sigma + {\cal P}\geq 0$.

From equations (\ref{surfenergy})-(\ref{surfpressure}), we deduce
\begin{equation} \label{sigma+P}
\sigma +{\cal P}=\frac{1}{8\pi a}
\left[(2-\zeta)\,\sqrt{1-\frac{b(a)}{a}} -
\frac{1-\frac{3M}{a}}{\sqrt{1-\frac{2M}{a}-\frac{\Lambda}{3}a^2}}
\right]  .
\end{equation}
In the next sections we shall find domains in which the NEC is
satisfied, by imposing that the surface energy density is
non-negative, $\sigma \geq 0$, i.e., $\sqrt{1-b(a)/a} \geq
\sqrt{1-2M/a-\Lambda a^2/3}$. A summary of the parameter domain
for which the null energy condition is satisfied, for all the
cases analyzed, is presented in Table \ref{t1}.

\subsubsection{Schwarzschild solution.}

Consider the Schwarzschild solution, $\Lambda=0$. We are
interested in finding the regions in which the WEC and the NEC at
the junction are satisfied, by imposing a non-negative surface
energy density, $\sigma \geq 0$, i.e., $\sqrt{1-b(a)/a} \geq
\sqrt{1-2M/a}$. For the particular case of $\zeta \leq 1$, from
equation (\ref{sigma+P}) we verify that $\sigma+{\cal P}\geq 0$ is
readily satisfied for $\forall \,a$.

For $1<\zeta <2$, the NEC is verified in the following region
\begin{equation}\label{Schwregion}
2M<a \leq 2M\,\left(\frac{\zeta-\frac{1}{2}}{\zeta-1}\right).
\end{equation}
For convenience, by defining a new parameter $\xi=2M/a$, equation
(\ref{Schwregion}) takes the form
\begin{equation}\label{Schwregion2}
\frac{\zeta-1}{\zeta-\frac{1}{2}} \leq \xi <1 \,.
\end{equation}

For $\zeta=2$, the NEC is satisfied for $\xi \geq 2/3$, i.e., $a
\leq 3M$. For $\zeta > 2$, we need to impose the NEC in the region
of equation (\ref{Schwregion2}); with $\sigma+{\cal P}<0$ for
$\xi<(\zeta-1)/(\zeta-1/2)$. See Table \ref{t1} for a summary of
the cases analyzed.

\subsubsection{Schwarzschild-de Sitter solution.}

For the Schwarzschild-de Sitter spacetime, $\Lambda > 0$, we shall
once again impose a non-negative surface energy density, $\sigma
\geq 0$. Consider the definitions $\beta=9 \Lambda M^2$ and
$\xi=2M/a$.

\begin{figure}[h]
  \centering
  \includegraphics[width=2.2in]{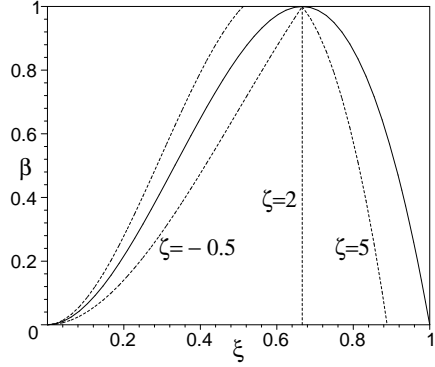}
  \caption{Analysis of the null energy condition for the
  Schwarzschild-de Sitter spacetime. We have considered the
  definitions $\beta=9 \Lambda M^2$ and $\xi=2M/a$.
  Only the region below the solid line is of interest. We have
  considered specific examples, and the NEC is
  obeyed to the right of each respective dashed curves,
  $\zeta=-0.5$,  $\zeta=2$ and $\zeta=5$.
  See text for details.}\label{fig4}
\end{figure}

For $\zeta < 2$ the condition $\sigma+{\cal P}\geq 0$ is readily
met for $\beta \leq \beta_0$, with $\beta_0$ given by
\begin{equation}\label{SdSbeta0}
\beta_0=\frac{27}{4}\,\frac{\xi^2}{(2-\zeta)}\,\left[(1-\zeta)
+\left(\zeta-\frac{1}{2}\right)\xi \right].
\end{equation}
Choosing a particular example, for instance $\zeta=-0.5$, consider
figure \ref{fig4}. The region of interest is shown below the solid
line, which is given by $\beta_r=27\xi^2(1-\xi)/4$. The case of
$\zeta=-0.5$ is depicted as a dashed curve, and the NEC is obeyed
to the right of the latter.

For $\zeta=2$, then the NEC is verified for $\forall \,\beta$ and
$\xi \geq 2/3$, i.e., $r_b<a \leq 3M$, with $r_b$ given by
equation (\ref{root1}). This analysis is depicted in figure
\ref{fig4}, with the NEC being satisfied to the right of the
dashed curve, represented by $\zeta=2$.

For the case of $\zeta>2$, the condition $\sigma+{\cal P}\geq 0$
needs to be imposed in the region $\beta_0 \leq \beta \leq
\beta_r$; and $\sigma+{\cal P}< 0$ for $\beta < \beta_0$. The
specific case of $\zeta=5$ is depicted as a dashed curve in figure
\ref{fig4}. The NEC needs to be imposed to the right of the
respective curve. See Table \ref{t1} for a summary of the cases
analyzed.

\subsubsection{Schwarzschild-anti de Sitter solution.}

Considering the Schwarzschild-anti de Sitter spacetime, $\Lambda <
0$, once again a non-negative surface energy density, $\sigma \geq
0$, is imposed. Consider the definitions $\gamma=9 |\Lambda| M^2$
and $\xi=2M/a$.

\begin{figure}[h]
  \centering
  \includegraphics[width=2.2in]{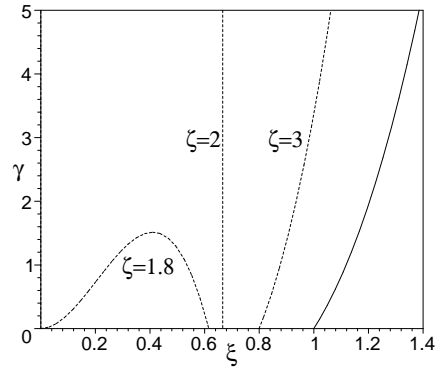}
  \caption{Analysis of the null energy condition for the
  Schwarzschild-anti de Sitter spacetime. We have considered the
  definitions $\gamma=9 |\Lambda| M^2$ and $\xi=2M/a$.
  The only area of interest is depicted to the left of the solid curve,
  given by $\gamma_r=27\xi^2(\xi-1)/4$. For the specific case of
  $\zeta=1.8$, the NEC is obeyed above the respective curve. For
  the cases of $\zeta=2$ and $\zeta=3$, the NEC is verified to the
  right of the respective dashed curves, and to the left of the
  solid line. See text for details.}\label{fig5}
\end{figure}

For $\zeta \leq 1$ the condition $\sigma+{\cal P}\geq 0$ is
readily met for $\forall \,\gamma$ and $\forall \,\xi$. For
$1<\zeta<2$, the NEC is satisfied in the region $\gamma \geq
\gamma_0$, with $\gamma_0$ given by
\begin{equation}\label{SadSbeta0}
\gamma_0=\frac{27}{4}\,\frac{\xi^2}{(2-\zeta)}\,\left[(\zeta-1)
-\left(\zeta-\frac{1}{2} \right)\xi  \right].
\end{equation}
The particular case of $\zeta=1.8$ is depicted in figure
\ref{fig5}. The region of interest is delimited  by the $\xi$-axis
and the area to the left of the solid curve, which is given by
$\gamma_r=27\xi^2(\xi-1)/4$. Thus, the NEC is obeyed above the
dashed curve represented by the value $\zeta=1.8$.

For $\zeta=2$, then $\sigma+{\cal P}\geq 0$ is verified for
$\forall \gamma$ and $\xi \geq 3/2$, i.e., $r_b<a \leq 3M$, with
$r_b$ given by equation (\ref{adsbhole}). Therefore, the NEC is
obeyed to the right of the dashed curve represented by $\zeta=2$,
and to the left of the solid line, $\gamma_r$.

For the case of $\zeta>2$, the condition $\sigma+{\cal P}\geq 0$
needs to be imposed in the region $\gamma_r \leq \gamma \leq
\gamma_0$. The specific case of $\zeta=3$ is depicted in figure
\ref{fig5} as a dashed curve. Thus, the NEC needs to be imposed in
the region to the right of the respective dashed curve and to the
left of the solid line, $\gamma_r$. See Table \ref{t1} for a
summary of the cases analyzed.

\bigskip

\begin{table}[h]
\begin{center}
\begin{tabular}[l]{|c|c|c|}
\hline
$\Lambda =0$ & $\zeta \leq 1$ & $\forall \,\xi$  \\
             & $\zeta > 1$    & $(\zeta-1)/(\zeta-1/2)\leq \xi < 1$  \\
\hline
             & $\zeta < 2$ & $\beta \leq \beta_0$ \\
$\Lambda >0$ & $\zeta = 2$ & $\forall \,\beta$ \quad and \quad $\xi \geq 2/3$ \\
             & $\zeta > 2$ & $\beta_0 \leq \beta \leq \beta_r$  \\
\hline
             & $\zeta \leq 1$  & $\forall \,\gamma$ \quad and \quad $\forall \,\xi$ \\
$\Lambda <0$ & $1<\zeta <2$ & $\gamma \geq \gamma_0$ \\
             & $\zeta = 2$  & $\forall \,\gamma$ \quad and \quad $\xi \geq 3/2$ \\
             & $\zeta > 2$  & $\gamma_r \leq \gamma \leq \gamma_0$  \\
 \hline
\end{tabular}
\caption{Parameter domain for which the null energy condition is
satisfied, i.e., $\sigma +{\cal P}\geq 0$, by imposing a positive
surface energy density, i.e., $\sigma\geq0$. We have defined the
parameters $\xi=2M/a$, $\beta=9 \Lambda M^2$ and $\gamma=9
|\Lambda| M^2$; $\beta_0$ and $\gamma_0$ are given by equation
(\ref{SdSbeta0}) and equation (\ref{SadSbeta0}), respectively. See
text for details.}\label{t1}
\end{center}
\end{table}

\subsection{Physical properties and characteristics of the surface stresses}

In this section we shall consider some interesting physical
properties and characteristics of the surface stresses for generic
wormholes by eliminating the form function $b(r)$, and writing
${\cal P}$ as a function of $\sigma$. Thus, taking into account
equations (\ref{surfenergy})-(\ref{surfpressure}), one may express
${\cal P}$ as a function of $\sigma$ by the following relationship
\begin{equation}
{\cal P}=\frac{1}{8\pi a}
\,\left[\frac{(1-\zeta)+(\zeta-\frac{1}{2})\,\frac{2M}{a}
-(2-\zeta)\frac{\Lambda}{3}a^2}{\sqrt{1-\frac{2M}{a}-
\frac{\Lambda}{3}a^2}} -4\pi a \zeta \sigma \right] .
        \label{Pfunctionsigma}
\end{equation}
We shall analyze equation (\ref{Pfunctionsigma}), namely, find
domains in which ${\cal P}$ assumes the nature of a tangential
surface pressure, ${\cal P}>0$, or a tangential surface tension,
${\cal P}<0$, for the Schwarzschild case, $\Lambda=0$, the
Schwarzschild-de Sitter spacetime, $\Lambda>0$, and for the
Schwarzschild-anti de Sitter solution, $\Lambda<0$. In the
analysis that follows we shall consider that $M$ is positive,
$M>0$.

\subsubsection{Schwarzschild spacetime.}

For the Schwarzschild spacetime, $\Lambda=0$, equation
(\ref{Pfunctionsigma}) reduces to
\begin{equation}
{\cal P}=\frac{1}{8\pi a}
\,\left[\frac{(1-\zeta)+(\zeta-\frac{1}{2})\,\frac{2M}{a}}{\sqrt{1-\frac{2M}{a}}}
-4\pi a \zeta \sigma \right] \,.
        \label{SchwarzPfunctionsigma}
\end{equation}
To find domains in which ${\cal P}$ is a tangential surface
pressure, ${\cal P}>0$, or a tangential surface tension, ${\cal
P}<0$, it is convenient to express equation
(\ref{SchwarzPfunctionsigma}) in the following compact form
\begin{equation}
{\cal P}=\frac{1}{16\pi M} \,
\frac{\Gamma(\xi,\zeta,\mu)}{\sqrt{1-\xi}}\,,
\label{SchwarzcompactP}
\end{equation}
with the dimensionless parameters given by $\xi=2M/a$ and
$\mu=8\pi M\sigma$. $\Gamma(\xi,\zeta,\mu)$ is defined as
\begin{eqnarray}\label{SchwarzGamma}
\Gamma(\xi,\zeta,\mu)=(1-\zeta)\,\xi+\left(\zeta-\frac{1}{2}\right)\xi^2-\mu
\zeta \sqrt{1-\xi}    \;.
\end{eqnarray}
One may now fix one or several of the parameters and analyze the
sign of $\Gamma(\xi,\zeta,\mu)$, and consequently the sign of
${\cal P}$.

\bigskip

{\it a. Fixed $\zeta$, varying $\xi$ and $\mu$.}

In this section we shall analyze the specific case of a fixed
value of $a\Phi'(a)$ and vary the values of the junction radius,
$a$, and of the surface energy density, $\sigma$, i.e., consider a
fixed value of the redshift parameter $\zeta$, varying the
parameters $(\xi,\mu)$. To analyze the sign of ${\cal P}$, it is
useful to consider a null tangential surface pressure, ${\cal
P}=0$, i.e., $\Gamma(\xi,\zeta,\mu)=0$. Thus, from equation
(\ref{SchwarzGamma}) we have
\begin{equation}\label{muzero}
\mu_0=\frac{(1-\zeta)\xi+(\zeta-1/2)\xi^2}{\zeta \sqrt{1-\xi}} \,,
\end{equation}
with $\zeta \neq 0$. It is necessary to separate the cases of
$\zeta=0$, $\zeta>0$ and $\zeta<0$, respectively.

Firstly, for the case of $\zeta=0$, equation (\ref{SchwarzGamma})
reduces to $\Gamma(\xi,\zeta=0,\mu)=\xi-\xi^2/2$, which is always
positive, as $0<\xi<1$, implying a tangential surface pressure,
${\cal P}>0$.

\begin{figure}[h]
  \centering
  \includegraphics[width=3.0in]{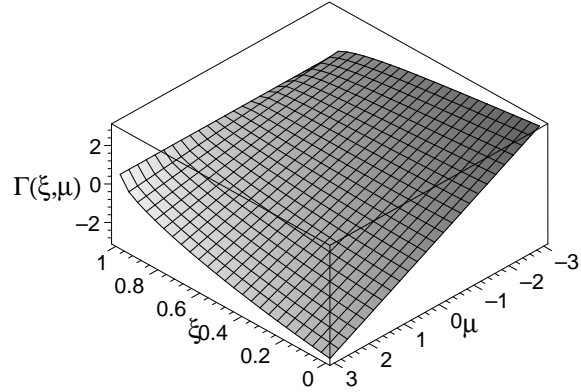}
  \caption{The surface represents the sign of ${\cal P}$ for the
  Schwarzschild spacetime with a
  constant redshift function, $\Phi'(r)=0$, i.e., $\zeta=1$.
  For non-positive values of $\mu$ and $\forall \,\xi$,
  we have a surface tangential
  pressure, ${\cal P}>0$. For extremely high values of $\xi$
  (close to the black hole event horizon) and $\forall \,\mu$,
  a surface pressure is also required to hold the structure
  against collapse. For high values of $\mu$ and low values of $\xi$,
  we have a tangential surface tension, ${\cal P}<0$, to hold the
  structure against expansion. See text for details.}\label{fig6}
\end{figure}

Secondly, for $\zeta>0$, a tangential surface pressure, ${\cal
P}>0$, is provided for $\mu<\mu_0$, and a tangential surface
tension, ${\cal P}<0$, for $\mu>\mu_0$. Note that a surface
boundary, with ${\cal P}=0$ and $\sigma=0$, is given by
$\xi=(\zeta-1)/(\zeta-1/2)$, for $\zeta>1$. For $0< \zeta \leq 1$,
we have a positive surface energy density, $\mu_0>0$, thus
satisfying the energy conditions, which is consistent with the
results of the previous section. The qualitative behavior for
$\zeta>0$ can be represented by the specific case of $\zeta=1$,
corresponding to a constant redshift function, depicted in figure
\ref{fig6}. For non-positive values of $\mu$ and $\forall \,\xi$ a
surface pressure, ${\cal P}>0$, is required to hold the thin shell
structure against collapse. Close to the black hole event horizon,
$a \rightarrow 2M$, i.e. $\xi \rightarrow 1$, a surface pressure
is also needed to hold the structure against collapse. For high
values of $\mu$ and low values of $\xi$, a surface tangential
tension, ${\cal P}<0$, is needed to hold the structure against
expansion. In particular, for the constant redshift function,
$\Phi'(r)=0$, and a null surface energy density, $\sigma=0$, i.e.,
$\zeta=1$ and $\mu=0$, respectively, equation (\ref{SchwarzGamma})
reduces to $\Gamma(\xi)=\xi^2/2$, from which we readily conclude
that ${\cal P}$ is non-negative everywhere, tending to zero at
infinity, i.e., $\xi \rightarrow 0$. This is a particular case
analyzed in \cite{LLQ}.

\begin{table}[h]
\begin{center}
\begin{tabular}[l]{|c|c|}
\hline
$\xi =0$ & $\Gamma(\xi=0,\zeta,\mu)>0$ \quad for \quad $\forall \,\mu$  \\
\hline
$\xi >0$ & $\Gamma(\xi,\zeta,\mu)>0$ \quad for \quad $\mu <\mu_0$  \\
         & $\Gamma(\xi,\zeta,\mu)<0$ \quad for \quad $\mu >\mu_0$  \\
\hline
$\xi <0$ & $\Gamma(\xi,\zeta,\mu)>0$ \quad for \quad $\mu >\mu_0$  \\
         & $\Gamma(\xi,\zeta,\mu)<0$ \quad for \quad $\mu <\mu_0$  \\
 \hline
\end{tabular}
\caption{The parameter domain of the sign of ${\cal P}$ for the
Schwarzschild solution. ${\cal P}$ is a tangential surface
pressure if $\Gamma(\xi,\zeta,\mu)>0$, and a surface tension if
$\Gamma(\xi,\zeta,\mu)<0$. $\Gamma(\xi,\zeta,\mu)$ and $\mu_0$ are
given by equation (\ref{SchwarzGamma}) and equation
(\ref{muzero}), respectively.. See text for details.}\label{t2}
\end{center}
\end{table}

\begin{figure}[h]
  \centering
  \includegraphics[width=3.0in]{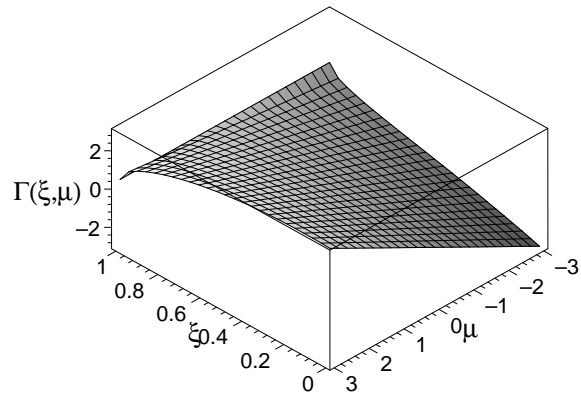}
  \caption{The surface is given by equation (\ref{SchwarzGamma}) for the
  Schwarzschild spacetime with $\zeta=-1$.
  For non-negative values of $\mu$ and $\forall \,\xi$,
  we have a surface tangential pressure, ${\cal P}>0$. For high
  values of $\xi$, i.e., close to the black hole event horizon,
  and for $\forall \,\mu$, a surface pressure is also required to
  hold the structure against collapse. For negative values of
  $\mu$ and low $\xi$, a tangential surface tension, ${\cal P}<0$,
  is required to hold the structure against expansion.
  See text for details.}\label{fig7}
\end{figure}

Finally, for the $\zeta<0$ case, we verify that a surface
pressure, ${\cal P}>0$, is obtained for $\mu>\mu_0$, and a
tangential surface tension, ${\cal P}<0$, for $\mu<\mu_0$,
contrary to the $\zeta>0$ analysis. The specific case of
$\zeta=-1$, depicted in figure \ref{fig7}, can be considered
representative for the qualitative behavior of $\zeta<0$. For
non-negative values of $\mu$ and for $\forall \,\xi$, a surface
pressure, ${\cal P}>0$, is required. Close to the black hole event
horizon, i.e., for high values of $\xi$, and for $\forall \,\mu$,
a surface pressure is also required to hold the structure against
collapse. For low negative values of $\mu$ and for low values of
$\xi$, a surface tension is needed, which is somewhat intuitive as
a negative surface energy density is gravitationally repulsive,
requiring a surface tension to hold the structure against
expansion. See Table \ref{t2} for a summary of the results
obtained.

\bigskip

{\it b. Fixed $\mu$, varying $\xi$, and $\zeta$.}

In this section we shall consider an alternative analysis. We
shall study the case of a fixed value for the surface energy
density, $\sigma$, and vary the values of the junction surface,
$a$, and of $a\Phi'(a)$, i.e., we fix the parameter $\mu$, varying
$\xi$ and $\zeta$. Equation (\ref{SchwarzGamma}) may be rewritten
in the form
\begin{eqnarray}\label{SchwarzGammamu}
\Gamma(\xi,\zeta,\mu)=\xi\,\left(1-\frac{\xi}{2}\right)-\sqrt{1-\xi}\left(\xi
\sqrt{1-\xi}-\mu \right)\zeta  \,.
\end{eqnarray}
Once again to analyze the sign of ${\cal P}$, we consider a null
tangential surface pressure, i.e., $\Gamma(\xi,\zeta,\mu)=0$.
Thus, from equation (\ref{SchwarzGammamu}) one finds the following
relationship
\begin{equation}\label{fixedmuzeta0}
\zeta_0=\frac{\xi(1-\xi/2)}{\sqrt{1-\xi}\left(\xi \,
\sqrt{1-\xi}-\mu \right)} \,,
\end{equation}
with $\mu \neq \xi \sqrt{1-\xi}$.

\begin{figure}[h]
  \centering
  \includegraphics[width=2.2in]{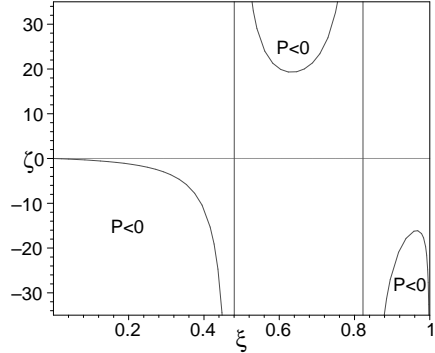}
  \caption{The curves are given by equation (\ref{fixedmuzeta0}) for
  the range $0<\mu<2\sqrt{3}/9$. We consider a particular example, i.e.,
  $\mu=1.7\sqrt{3}/9$. The area above the curve in the
  intermediate region, and below the curves in the two outer regions,
  correspond to surface tensions. The intermediate region expands
  outwards to the points
  $\xi \rightarrow 0$ and $\xi \rightarrow 1$, respectively, as
  $\mu \rightarrow 0$, and contracts to the
  point $\xi \rightarrow 2/3$, as $\mu \rightarrow 2\sqrt{3}/9$.
  See text for details.}\label{fig8}
\end{figure}

For the specific case of $\mu =\xi \sqrt{1-\xi}$, equation
(\ref{SchwarzGammamu}) reduces to
$\Gamma(\xi,\zeta,\mu)=\xi(1-\xi/2)$, which is always positive,
implying a surface pressure. For this case $\mu$ attains a maximum
at $\mu_{\rm max}=2\sqrt{3}/9$ for $\xi=2/3$, i.e., $a=3M$.
Therefore, for fixed values of $\mu$ in the interval
$0<\mu<2\sqrt{3}/9$, one has three regions to analyze the sign of
${\cal P}$. Consider, for simplicity, the specific case of
$\mu=1.7\sqrt{3}/9$, represented in figure \ref{fig8}. The regions
corresponding to the surface tensions are shown. As $\mu
\rightarrow 0$ the intermediate region expands outwards to the
points $\xi \rightarrow 0$ and $\xi \rightarrow 1$, respectively;
as $\mu \rightarrow 2\sqrt{3}/9$, the intermediate region
contracts to the point $\xi \rightarrow 2/3$.

\begin{figure}[h]
  \centering
  \includegraphics[width=2.2in]{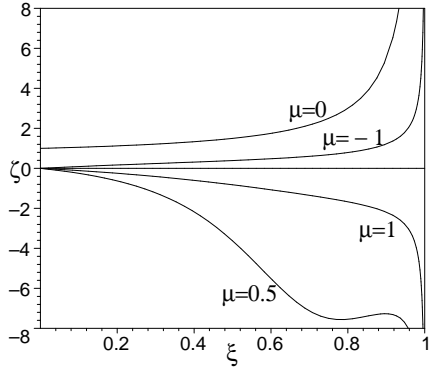}
  \caption{The curves are given by equation (\ref{fixedmuzeta0}).
  For the cases of $\mu=0$ and $\mu=-1$ represented,
  below the respective curves we have a surface pressure, ${\cal
  P}>0$, and above the curves, a surface tension, ${\cal P}<0$.
  For the cases of $\mu=0.5$ and $\mu=1$, we have the inverse
  behavior, i.e., below the respective curves we have a surface
  tension, ${\cal P}<0$, and above the curves a surface pressure,
  ${\cal P}>0$. See text for details.}\label{fig9}
\end{figure}

For $\mu <\xi \sqrt{1-\xi}$, we have $\Gamma(\xi,\zeta,\mu)>0$,
implying a surface pressure, ${\cal P}>0$, for $\zeta<\zeta_0$;
and $\Gamma(\xi,\zeta,\mu)<0$, providing a surface tension, ${\cal
P}<0$, for $\zeta>\zeta_0$. The analysis for non-positive values
of the surface energy density, $\sigma \leq 0$, is fairly
straightforward. The qualitative behavior for $\mu \leq 0$ is
similar to that of the cases presented in figure \ref{fig9},
namely, that of $\mu=0$ and $\mu=-1$, respectively. Below the
respective curves, we have a surface pressure, ${\cal P}>0$, and
above the respective curves, we have a surface tension, ${\cal
P}<0$.

Finally, for $\mu >\xi \sqrt{1-\xi}$, one has a surface pressure,
${\cal P}>0$, for $\zeta>\zeta_0$; and a surface tension, ${\cal
P}<0$, for $\zeta<\zeta_0$. For values of $\mu
>2\sqrt{3}/9$, the qualitative behavior is similar to the cases of
$\mu=0.5$ and $\mu=1$, which are represented in figure \ref{fig9}.
Above the respective curves, we have a surface pressure, ${\cal
P}>0$, and below the curves, a surface tension, ${\cal P}<0$.

{\it c. Fixed $\xi$, varying $\zeta$ and $\mu$.}

For an alternative analysis consider a fixed value for the
junction radius, $a$, varying $a\Phi'(a)$ and the surface energy
density, $\sigma$, i.e.,  we fix $\xi$ and vary the parameters
$\zeta$ and $\mu$. Equation (\ref{SchwarzGamma}) with $\xi=0$,
reduces to $\Gamma(\xi=0,\zeta,\mu)=-\mu \,\zeta$. Qualitatively,
we verify that for either positive values for both $\zeta$ and
$\mu$, or for negative values for both $\zeta$ and $\mu$, we have
$\Gamma(\xi=0,\zeta,\mu)<0$, implying a surface tension. If
$\zeta$ is negative and $\mu$ positive, or vice-versa, we verify
that $\Gamma(\xi=0,\zeta,\mu)>0$, implying a tangential surface
pressure. Now, as $\xi$ increases from $0$ to $1$,
$\Gamma(\zeta,\mu)$ attains the value of a constant surface
$\Gamma(\xi=1,\zeta,\mu)\rightarrow 1/2$, as $\xi \rightarrow 1$,
i.e., $a \rightarrow 2M$, is reached.

\subsubsection{Schwarzschild-de Sitter spacetime.}

For the Schwarzschild-de Sitter spacetime with $\Lambda>0$, to
analyze the sign of ${\cal P}$, it is convenient to express
equation (\ref{Pfunctionsigma}) in the following compact form
\begin{equation}
{\cal P}=\frac{1}{16\pi M} \,
\frac{\Gamma(\xi,\zeta,\mu,\beta)}{\sqrt{1-\xi-\frac{4\beta}{27\xi^2}}}\,,
\label{compactP}
\end{equation}
with $\xi=2M/a$, $\mu=8\pi M\sigma$ and $\beta=9\Lambda M^2$.
$\Gamma(\xi,\zeta,\mu,\beta)$ is defined as
\begin{eqnarray}\label{GammadS}
\Gamma(\xi,\zeta,\mu,\beta)&=&(1-\zeta)\,\xi+\left(\zeta-\frac{1}{2}\right)\xi^2
-\frac{4\beta}{27\xi}\,(2-\zeta)
     \nonumber   \\
&& -\mu \zeta \sqrt{1-\xi-\frac{4\beta}{27\xi^2}}    \;.
\end{eqnarray}
To analyze the sign of $\Gamma(\xi,\zeta,\mu,\beta)$, and
consequently the sign of ${\cal P}$, one may fix several of the
parameters.


{\it a. Null surface energy density.}

Consider a null surface energy density, $\sigma=0$, i.e., $\mu=0$.
Thus, equation (\ref{GammadS}) reduces to
\begin{equation}\label{GammasigmadS}
\Gamma(\xi,\zeta,\beta)=(1-\zeta)\,\xi+\left(\zeta-\frac{1}{2}\right)\,\xi^2-
(2-\zeta)\,\frac{4\beta}{27\xi}\,.
\end{equation}
To analyze the sign of ${\cal P}$, we shall consider a null
tangential surface pressure, i.e., $\Gamma(\xi,\zeta,\beta)=0$, so
that from equation (\ref{GammasigmadS}) we have the following
relationship
\begin{equation}\label{beta0}
\beta_0=\frac{27}{4}\frac{\xi^2}{(2-\zeta)}\left[(1-\zeta)
+\left(\zeta-\frac{1}{2}\right)\,\xi \right]  \,,
\end{equation}
with $\zeta \neq 2$, which is identical to equation
(\ref{SdSbeta0}).

For the particular case of $\zeta=2$, from equation
(\ref{GammasigmadS}), we have
$\Gamma(\xi,\zeta=2,\beta)=\xi(3\xi/2-1)$. A surface boundary,
$\Gamma(\xi,\zeta=2,\beta)=0$, is presented at $\xi=2/3$, i.e.,
$a=3M$. A surface pressure, $\Gamma(\xi,\zeta=2,\beta)>0$, is
given for $\xi>2/3$, i.e., $r_b<a<3M$, and a surface tension,
$\Gamma(\xi,\zeta=2,\beta)<0$, for $\xi<2/3$, i.e., $3M<a<r_c$.

For $\zeta<2$, from equation (\ref{GammasigmadS}), a surface
pressure, $\Gamma(\xi,\zeta,\beta)>0$, is met for $\beta<\beta_0$,
and a surface tension, $\Gamma(\xi,\zeta,\beta)<0$, for
$\beta>\beta_0$. The specific case of a constant redshift
function, i.e., $\zeta=1$, is analyzed in \cite{LLQ} (For this
case, equation (\ref{beta0}) is reduced to $\beta_0=27\xi^3/8$, or
$M=\Lambda a^3/3$). For the qualitative behavior, the reader is
referred to the particular case of $\zeta=-0.5$ provided in figure
\ref{fig4}. To the right of the curve a surface pressure, ${\cal
P}>0$, is given and to the left of the respective curve a surface
tension, ${\cal P}<0$. One verifies that as $\zeta \rightarrow
-\infty$, the curve superimposes with the solid line, $\beta_r$,
and one only has a surface pressure for the entire region of
interest.

For $\zeta>2$, from equation (\ref{GammasigmadS}), a surface
pressure, $\Gamma(\xi,\zeta,\beta)>0$, is given for
$\beta_0<\beta<\beta_r$, and a surface tension,
$\Gamma(\xi,\zeta,\beta)<0$, for $\beta<\beta_0$. Once again the
reader is referred to figure \ref{fig4} for a qualitative analysis
of the behavior for the particular case of $\zeta=5$. A surface
pressure is given to the right of the curve and a surface tension
to the left. For $\zeta \rightarrow +\infty$, the curve
superimposes with the solid line, $\beta_r$, and a surface tension
is given for the entire region of interest.

Note that for the analysis considered in this section, namely, for
a null surface energy density, the WEC, and consequently the NEC,
are satisfied only if ${\cal P} \geq 0$. The results obtained are
consistent with those of the section regarding the energy
conditions at the junction surface, for the Schwarzschild-de
Sitter spacetime, considered above. See Table \ref{t3} for a
summary of the results obtained.


\begin{table}[h]
\begin{center}
\begin{tabular}[l]{|c|c|}
\hline
$\zeta <2$ & $\Gamma(\xi,\zeta,\beta)>0$, \quad $\beta <\beta_0$  \\
         & $\Gamma(\xi,\zeta,\beta)<0$, \quad $\beta >\beta_0$  \\
\hline
         & $\Gamma(\xi,\zeta=2,\beta)=0$, \quad $\forall \,\beta$ \quad and \quad $\xi=2/3$  \\
$\zeta =2$ & $\Gamma(\xi,\zeta=2,\beta)>0$, \quad $\forall \,\beta$ \quad and \quad $\xi>2/3$   \\
         & $\Gamma(\xi,\zeta=2,\beta)<0$, \quad $\forall \,\beta$ \quad and \quad $\xi<2/3$   \\
\hline
$\zeta >2$ & $\Gamma(\xi,\zeta,\beta)>0$, \quad $\beta_0 <\beta<\beta_r$  \\
         & \hspace{-0.95cm}$\Gamma(\xi,\zeta,\beta)<0$, \quad $\beta <\beta_0$  \\
 \hline
\end{tabular}
\caption{The parameter domain of the sign of ${\cal P}$, for the
Schwarzschild-de Sitter solution, $\Lambda>0$. Considering the
particular case of a null surface energy density, i.e., $\mu=0$,
${\cal P}$ is a tangential surface pressure if
$\Gamma(\xi,\zeta,\beta)>0$, and a surface tension if
$\Gamma(\xi,\zeta,\beta)<0$. $\Gamma(\xi,\zeta,\beta)$ and
$\beta_0$ are given by equation (\ref{GammasigmadS}) and equation
(\ref{beta0}), respectively. See text for details.}\label{t3}
\end{center}
\end{table}



{\it b. Constant redshift function.}

We shall next consider a constant redshift function, $\Phi'(r)=0$,
i.e., $\zeta=1$. Thus equation (\ref{GammadS}) is reduced to
\begin{equation}\label{GammadSzetaunity}
\Gamma(\xi,\mu,\beta)=\frac{\xi^2}{2} -\frac{4\beta}{27\xi} -\mu
\, \sqrt{1-\xi-\frac{4\beta}{27\xi^2}}    \,.
\end{equation}
Considering a null tangential surface pressure,
$\Gamma(\xi,\mu,\beta)=0$, equation (\ref{GammadSzetaunity}) takes
the following form
\begin{equation}
\mu_0=\frac{\frac{\xi^2}{2}
-\frac{4\beta}{27\xi}}{\sqrt{1-\xi-\frac{4\beta}{27\xi^2}}}  \,.
\end{equation}
We verify that a surface pressure, ${\cal P}>0$, is given for
$\mu<\mu_0$, and a surface tension, ${\cal P}<0$, for $\mu>\mu_0$.
A surface boundary, $\mu_0=0$, is verified for
$\beta_0=27\xi^3/8$.

\subsubsection{Schwarzschild-anti de Sitter spacetime.}

For the Schwarzschild-anti de Sitter spacetime, $\Lambda<0$, to
analyze the sign of ${\cal P}$, equation (\ref{Pfunctionsigma}) is
expressed as
\begin{equation}
{\cal P}=\frac{1}{16\pi M} \,
\frac{\Gamma(\xi,\zeta,\mu,\gamma)}{\sqrt{1-\xi+\frac{4\gamma}{27\xi^2}}}\,,
\label{SadScompactP}
\end{equation}
with the parameters given by $\xi=2M/a$, $\mu=8\pi M\sigma$ and
$\gamma=9|\Lambda| M^2$, respectively.
$\Gamma(\xi,\zeta,\mu,\gamma)$ is defined as
\begin{eqnarray}\label{GammaSadS}
\Gamma(\xi,\zeta,\mu,\gamma)&=&(1-\zeta)\,\xi+\left(\zeta-\frac{1}{2}\right)\xi^2
+\frac{4\gamma}{27\xi}\,(2-\zeta)
     \nonumber   \\
&& -\mu \zeta \sqrt{1-\xi+\frac{4\gamma}{27\xi^2}}    \;.
\end{eqnarray}
As in the Schwarzschild-de Sitter solution we shall analyze the
cases of a null surface energy density, $\sigma=0$, and a constant
redshift function, $\Phi'(r)=0$, i.e., $\mu=0$ and $\zeta=1$,
respectively.


{\it a. Null surface energy density.}

For a null surface energy density, $\sigma=0$, i.e., $\mu=0$,
equation (\ref{GammaSadS}) reduces to
\begin{equation}\label{GammasigmaSadS}
\Gamma(\xi,\zeta,\gamma)=(1-\zeta)\,\xi+\left(\zeta-\frac{1}{2}\right)\,\xi^2+
(2-\zeta)\,\frac{4\gamma}{27\xi}\,,
\end{equation}
To analyze the sign of ${\cal P}$, once again we shall consider a
null tangential surface pressure, i.e.,
$\Gamma(\xi,\zeta,\gamma)=0$, so that from equation
(\ref{GammasigmaSadS}) we have
\begin{equation}\label{SadSbeta}
\gamma_0=\frac{27}{4}\frac{\xi^2}{(2-\zeta)}\left[(\zeta-1)
-\left(\zeta-\frac{1}{2}\right)\,\xi \right]  \,,
\end{equation}
with $\zeta \neq 2$, which is identical to equation
(\ref{SadSbeta0}).

For the particular case of $\zeta=2$, from equation
(\ref{GammasigmaSadS}), we have
$\Gamma(\xi,\zeta=2,\gamma)=\xi(3\xi/2-1)$, which is null at
$\xi=2/3$, i.e., $a=3M$. A surface pressure,
$\Gamma(\xi,\zeta=2,\gamma)>0$, is given for $\xi>2/3$, i.e.,
$r_b<a<3M$, and a surface tension, $\Gamma(\xi,\zeta=2,\gamma)<0$,
for $\xi<2/3$, i.e., $a>3M$. The reader is referred to the
particular case of $\zeta=2$, depicted in figure \ref{fig5}. A
surface pressure is given to the right of the respective dashed
curve, and a surface tension to the left.

For $\zeta \leq 1$, a surface pressure,
$\Gamma(\xi,\zeta,\gamma)>0$, is given for $\forall \; \gamma$ and
$\forall \;\xi$. For $1<\zeta<2$, a surface pressure,
$\Gamma(\xi,\zeta,\gamma)>0$, is given for $\gamma>\gamma_0>0$;
and a surface tension, $\Gamma(\xi,\zeta,\gamma)<0$, is provided
for $0<\gamma<\gamma_0$. The particular case of $\zeta=1.8$ is
depicted in figure \ref{fig5}, in which a surface pressure is
presented above the respective dashed curve, and a surface tension
is presented in the region delimited by the curve and the
$\xi$-axis.

For $\zeta>2$, a surface pressure, $\Gamma(\xi,\zeta,\gamma)>0$,
is met for $\gamma_r<\gamma<\gamma_0$, and a surface tension,
$\Gamma(\xi,\zeta,\gamma)<0$, for $\gamma>\gamma_0$. The specific
case for $\zeta=3$ is depicted in figure \ref{fig5}. A surface
pressure is presented to the right of the respective curve, and a
surface tension to the left.

Once again, the analysis considered in this section is consistent
with the results obtained in the section regarding the energy
conditions at the junction surface, for the Schwarzschild-anti de
Sitter spacetime, considered above. This is due to the fact that
for the specific case of a null surface energy density, the
regions in which the WEC and NEC are satisfied coincide with the
range of ${\cal P} \geq 0$. See Table \ref{t4} for a summary of
the results obtained.


\begin{table}[h]
\begin{center}
\begin{tabular}[l]{|c|c|}
\hline
$\zeta \leq 1$ & $\Gamma(\xi,\zeta,\gamma)>0$,  \quad $\forall \,\gamma$ \quad and \quad $\forall \,\xi$ \\
\hline
$1<\zeta <2$ & $\Gamma(\xi,\zeta,\gamma)>0$, \quad $\gamma >\gamma_0>0$  \\
         & $\Gamma(\xi,\zeta,\gamma)<0$, \quad  $0<\gamma <\gamma_0$  \\
\hline
         & $\Gamma(\xi,\zeta=2,\gamma)=0$, \quad $\forall \,\gamma$ \quad and \quad $\xi=2/3$  \\
$\zeta =2$ & $\Gamma(\xi,\zeta=2,\gamma)>0$, \quad $\forall \,\gamma$ \quad and \quad $\xi>2/3$   \\
         & $\Gamma(\xi,\zeta=2,\gamma)<0$, \quad $\forall \,\gamma$ \quad and \quad $\xi<2/3$   \\
\hline
$\zeta >2$ & $\Gamma(\xi,\zeta,\gamma)>0$, \quad $\gamma_r <\gamma<\gamma_0$  \\
         & \hspace{-0.94cm}$\Gamma(\xi,\zeta,\gamma)<0$, \quad $\gamma >\gamma_0$  \\
 \hline
\end{tabular}
\caption{The parameter domain for the sign of ${\cal P}$,
considering the Schwarzschild-anti de Sitter solution,
$\Lambda<0$. Considering the particular case of a null surface
energy density, i.e., $\mu=0$, ${\cal P}$ is a tangential surface
pressure if $\Gamma(\xi,\zeta,\gamma)>0$, and a surface tension if
$\Gamma(\xi,\zeta,\gamma)<0$. $\Gamma(\xi,\zeta,\gamma)$ and
$\gamma_0$ are given by equation (\ref{GammasigmaSadS}) and
equation (\ref{SadSbeta}), respectively. See text for
details.}\label{t4}
\end{center}
\end{table}



{\it b. Constant redshift function.}

For the case of a constant redshift function, $\Phi'(r)=0$, i.e.,
$\zeta=1$, equation (\ref{GammaSadS}) is reduced to
\begin{eqnarray}\label{GammaSadSzetaunity}
\Gamma(\xi,\mu,\gamma)=\frac{\xi^2}{2} +\frac{4\gamma}{27\xi} -\mu
\, \sqrt{1-\xi+\frac{4\gamma}{27\xi^2}}    \,.
\end{eqnarray}
To analyze the sign of ${\cal P}$, consider
$\Gamma(\xi,\mu,\gamma)=0$, so that equation
(\ref{GammaSadSzetaunity}) takes the following form
\begin{equation}
\mu_0=\frac{\frac{\xi^2}{2}
+\frac{4\gamma}{27\xi}}{\sqrt{1-\xi+\frac{4\gamma}{27\xi^2}}} \,,
\end{equation}
which is always positive. A surface pressure, ${\cal P}>0$, is
given for $0<\mu<\mu_0$, and a surface tension, ${\cal P}<0$, for
$\mu>\mu_0$.

\subsection{Pressure balance equation}

One may obtain an equation governing the behavior of the radial
pressure in terms of the surface stresses at the junction boundary
from the following identity \cite{Visser,Musgrave}
\begin{equation}\label{generalpressurebalance}
\left[\,T^{\rm
total}_{\hat{\mu}\hat{\nu}}\,n^{\hat{\mu}}n^{\hat{\nu}}
\right]=\frac{1}{2}(K^{i\;+}_{\;\,j} +
K^{i\;-}_{\;\,j})\,S^{j}_{\;\,i}   \,,
\end{equation}
where $T^{\rm
total}_{\hat{\mu}\hat{\nu}}=T_{\hat{\mu}\hat{\nu}}-g_{\hat{\mu}\hat{\nu}}\,\Lambda/8\pi$
is the total stress-energy tensor, and the square brackets denotes
the discontinuity across the thin shell, i.e.,
$[X]=X^{+}|_{\Sigma}-X^{-}|_{\Sigma}$. Taking into account the
values of the extrinsic curvatures, equations
(\ref{Kplustautau})-(\ref{Kminustheta}), and noting that the
tension acting on the shell is by definition the normal component
of the stress-energy tensor,
$-\tau=T_{\hat{\mu}\hat{\nu}}\,n^{\hat{\mu}}n^{\hat{\nu}}$, we
finally have the following pressure balance equation
\begin{eqnarray}\label{pressurebalance}
&&\fl \left(-\tau^+(a)-\frac{\Lambda ^+}{8\pi} \right) -
\left(-\tau ^-(a)-\frac{\Lambda ^-}{8\pi} \right)=
      \frac{1}{a}\,\left(\sqrt{1-\frac{2M}{a}-\frac{\Lambda}{3}a^2}
+\sqrt{1-\frac{b(a)}{a}}\;\right)\,{\cal P}
      \nonumber       \\
&&-\left(\frac{\frac{M}{a^2}- \frac{\Lambda}{3}a}
{\sqrt{1-\frac{2M}{a}-\frac{\Lambda}{3}a^2}}+\Phi'(a)\,\sqrt{1-\frac{b(a)}{a}}
\right) \frac{\sigma}{2} ,
\end{eqnarray}
where the $\pm$ superscripts correspond to the exterior and
interior spacetimes, respectively. Equation
(\ref{pressurebalance}) relates the difference of the radial
tension across the shell in terms of a combination of the surface
stresses, $\sigma$ and ${\cal P}$, given by equations
(\ref{surfenergy})-(\ref{surfpressure}), respectively, and the
geometrical quantities.

Note that for the exterior vacuum solution we have $\tau^+=0$. For
the particular case of a null surface energy density, $\sigma=0$,
and considering that the interior and exterior cosmological
constants are equal, $\Lambda^-=\Lambda^+$, equation
(\ref{pressurebalance}) reduces to
\begin{equation}
\tau
^-(a)=\frac{2}{a}\,\sqrt{1-\frac{2M}{a}-\frac{\Lambda}{3}a^2}\;\,{\cal
P} \,.
\end{equation}
For a radial tension, $\tau^-(a)>0$, acting on the shell from the
interior, a tangential surface pressure, ${\cal P}>0$, is needed
to hold the thin shell form collapsing. For a radial interior
pressure, $\tau^-(a)<0$, then a tangential surface tension, ${\cal
P}<0$, is needed to hold the structure form expansion.


\section{Conclusion}

We have constructed wormhole solutions by matching an interior
solution to a vacuum exterior spacetime, with a generic
cosmological constant, at a junction surface. In the spirit of
minimizing the usage of exotic matter, regions satisfying the weak
and null energy conditions at the junction surface were
determined. Thus, models of spherically symmetric traversable
wormholes were constructed where the null energy condition
violating matter is confined to the region $r_0\leq r <a$, while
{\it quasi-normal} matter resides at $a$ (here, {\it quasi-normal}
meaning matter that does not violate the weak and null energy
conditions; see \cite{VKD} for similar definitions).

Thin shells or domain walls in field theory arise in models with
spontaneously broken discrete symmetries \cite{Vilenkin}. The
model under consideration involves a set of real scalar fields
$\phi_i$ with a Lagrangian of the form ${\cal
L}=\frac{1}{2}\,(\partial_\mu \phi_i)^2-V(\phi)$, where the
potential $V(\phi)$ has a discrete set of degenerate minima. Thus,
one expects that for some chosen $\phi_i$ and $V(\phi)$, one gets
the thin shells that obey the energy conditions analyzed in this
paper. We have considered that the wormhole configurations studied
are {\it a priori} stable. However, one may analyze, for instance,
the dynamical stability of the thin shell, considering linearized
radial perturbations around stable solutions
\cite{LoboCraw-dynamic}. Estimates for the surface stresses,
considering a specific form function, were studied, and the
characteristics and several physical properties of the surface
stresses, for generic wormhole configurations, were explored,
namely, regions where the sign of the tangential surface pressure
is positive and negative (surface tension) were specified. An
equation governing the behavior of the radial pressure across the
junction surface was also deduced.


\ack The author acknowledges many fruitful and stimulating
discussions with Jos\'e P. S. Lemos, and thanks an anonymous
referee, whose comments led to an overall improvement of the
manuscript.


\end{document}